# Limiting Regimes for Electron-Beam Induced Deposition of Copper from Aqueous Solutions Containing Surfactants

*Samaneh Esfandiarpour, and J. Todd Hastings*

Electrical and Computer Engineering, University of Kentucky, Lexington, KY 40506

**ABSTRACT:** Focused electron beam induced deposition of pure materials from aqueous solutions has been of interest in recent years. However, controlling the liquid film in partial vacuum is challenging. Here we modify the substrate to increase control over the liquid layer in order to conduct a parametric study of copper deposition in an environmental SEM. We identified the transition from electron to mass-transport limited deposition as well as two additional regimes characterized by aggregated and high-aspect ratio deposits. We observe a high deposition efficiency of 1 to 10 copper atoms per primary electron that is consistent with a radiation chemical model of the deposition process.

## 1. Introduction

Focused electron beam induced processing (FEBIP) is a versatile technique that allows the direct deposition and etching of nanostructures. FEBIP provides a means of nanoscale, 3D rapid prototyping, and it has traditionally been conducted using gaseous precursors and etchants [1-10]. A detailed review of the precursors and their interactions with surfaces for these processes is reported in [11]. Now, liquid-phase focused electron beam induced processing (LP-FEBIP) is being investigated due to several potential benefits over gas phase techniques including material purity, process throughput, and access to new materials. In this method, deposition or etching occurs at the interface between a substrate and a bulk liquid. To date, Pt [12-14], Au [15], Ag [16, 17], amorphous carbon [18], Pd [19, 20], bimetallic alloys (AuAg and AuPt) [21], semiconductors (CdS) [22] have been successfully deposited in sealed liquid cells with an electron transparent membrane. Moreover, TEM liquid cells have been used to study LP-FEBIP, including bubble formation and nucleation and growth of gold [23], lead sulfide (PbS) nanoparticle growth [24], and electrochemical deposition of copper clusters [25] Dynamic phenomena at liquid−solid interfaces in electrode-electrolyte systems during charge and discharge processes have also been studied, along with nucleation, growth, and self-assembly of colloidal nanocrystals [26].

Recently, liquid-phase FEBID (LP-FEBID) on bulk substrates without liquid cells was accomplished using in-situ injection,[18, 27] environmental SEM (ESEM), or a combination of the two.[28, 29] Liquid film thickness is critical for understanding and controlling the LP-FEBIP. Specifically, pattern reproducibility and consistency as well as electron scattering events are affected directly by liquid thickness.[27- 31] However, controlling the thin liquid film in a partial vacuum has remained a serious challenge. Previously, we used shallow microwells to produce a meniscus which yielded a thin liquid layer close to the centre of the well.[29] We used this approach for the e-beam induced deposition of copper, a material that is difficult to deposit with high purity from gas-phase metalorganic precursors.[32-34] High purity and high-resolution copper deposition was readily obtained from the liquid precursor. However, control of the precursor concentration and the film thickness remained challenging due to the ultra-small liquid volume.

In this article, we demonstrate increased control over the liquid layer by creating deep micro-wells connected to a shallower well via a channel. The microcapillary effect of the channel results in dragging the liquid solution from the larger reservoir well into the other microwell. The second microwell and the channel have controlled depth suitable for patterning. In this way, we provide larger areas appropriate for patterning. Moreover, since the precursor in the channel is connected to the solution in the large well, we have a very good estimate of the concentration of the solution. We used these structures to determine the limiting regimes for copper deposition from aqueous solutions using focused electron beam.

## 2. Experimental Methods

All deposition experiments were carried out using an environmental scanning electron microscope (ESEM, Quanta 250 FEG, Thermo Fisher Scientific, Hillsboro, OR, USA). Patterning was controlled by a Raith Elphy 7 electron beam patterning system. A gaseous secondary electron detector (GSED) was used to collect secondary electron signals (SE) in the ESEM. A Peltier cooling stage was used to carefully control samples temperature and hence, the condensation and evaporation of water from the sulfuric acid solution on the surface. 30 kV acceleration voltage beam and a current of 700 pA were used for every experiment.

## 2.1 Microwells and Liquid Control

To carefully study the effect of $Cu^{2+}$ ions concentration and also to precisely control the liquid film in-situ, a modified substrate was used. The modified configuration consists of 50 μm deep reservoirs connected to shallower wells of 3 μm depth via a channel. Reservoir diameters ranged from 200 to 500 μm and channel widths ranged from 60 to 100 μm. Typical experiments used three microwells with the same orientation, but with different size reservoirs. This allowed us to run the deposition process in each well with a slight change (~ 0.1 Torr) in the chamber pressure. These structures were created by deep reactive-ion etching into the silicon wafer as shown in Figure 1(a). Before the deposition process, the substrates were rinsed with acetone, IPA and DI-$H_2O$ followed by the *in-situ* plasma cleaning to improve wetting. The microcapillary effect of the channel results in dragging the liquid solution from the larger reservoir well into the other microwell. By carefully controlling the substrate temperature and water-vapor pressure in the chamber we can reach equilibrium where the liquid film is stable. At this point, we expose the desired region in the channel or the second well to the electron beam in order to deposit copper patterns. Copper deposition from several liquid precursors was studied in [29]. It was concluded that an aqueous solution of copper sulfate and sulphuric acid resulted in high purity copper deposits, (≈95 at. %). The same precursor was used in this study as well.

The precursor was loaded into the 50 μm deep well *ex-situ* using a custom liquid dispensing system as illustrated in Figure 1(b). We filled up the deep reservoir to the nominal volume to be able to estimate the final concentration of copper ions during the deposition process. The dispensing system is controlled by a XP3000 syringe pump with a 50 μl syringe size and minimum flow rate of 0.0025 ml /min. the pump was controlled via a computer interface program to lower or raise the syringe with fine steps and controlled speed. 5-μm inner-diameter prepulled glass micro pipettes (TIP5TW1, World precision instruments.) were used to deliver the precursor into the microwells. A stereo microscope with a CMOS camera (Point Grey Research, Inc.) was used to observe the glass micro pipette position with respect to the reservoir. A three-axis stage under the microscope allowed us to precisely position the substrate. Z control axis was used to lower the stage as soon as the reservoir was filled completely. In case of overflow during filling, the substrate would be replaced so that there was no concern about estimating the concentration of copper precursor accurately.

The solution was further hydrated by controlling the temperature and pressure in the ESEM with a water vapor ambient. Figure 1(c)-(f) display the in-situ hydration of precursor on the bulk substrate schematically. It is seen that by cooling the substrate and increasing the water vapor pressure, in our case to 3°C and ~ 6 torr, a liquid film forms on the surface that can be used as the deposition medium containing the metal ions. The solution was rehydrated until the secondary-electron signal was constant across the reservoir. The absence of signal variation is indicative of a flat liquid film. The absence of a negative or positive meniscus indicated that the reservoir was neither under nor over filled, respectively.

As the liquid fills the deep well, it starts to flow through the channel toward the other well as shown in Figure 2. As mentioned earlier, the precursor in the channel is connected to the solution in the large well, so we have a very good estimate of the concentration of the solution. In some experiments, overflow of the reservoir occurred during initial rehydration and left precursor residue on the surrounding substrate. Such an event prevents accurate knowledge of the precursor concentration, and thus these experiments were stopped, and a fresh substrate was introduced. After stabilizing the liquid film, an electron beam is applied to the specific location with the predefined pattern. Then the substrate is reheated to the room temperature, the liquid film retracts from the channel, and the sample is removed from the chamber. The deposits are subsequently studied in the high vacuum mode of the microscope.

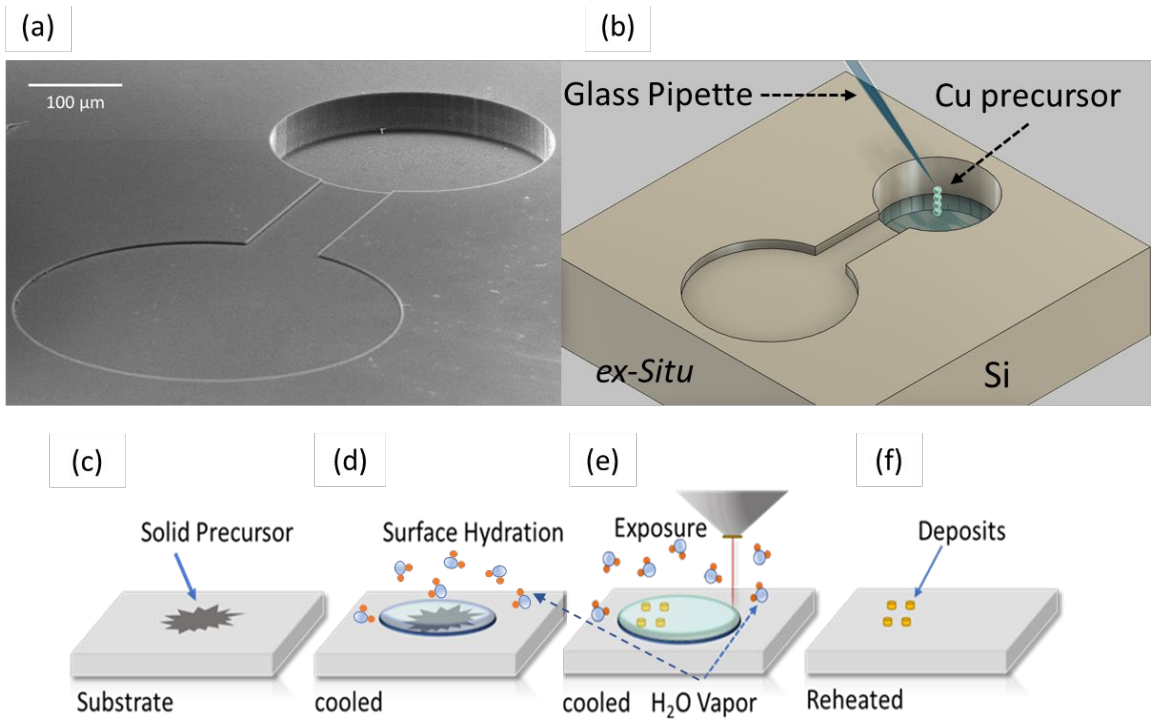

Figure 1: (a) 60° tilted view of deep reactive-ion etched microwells in the Si substrate. A 300 µm diameter reservoir is connected to a 400 µm diameter well via a 70 µm width and 100 µm length channel. The reservoir depth is 50 µm and the channel and well depths are 3 µm. (b) E*x-situ* dispensing of the copper precursor in the reservoir, (c-f) Schematic of electron-beam induced deposition from a bulk liquid in an environmental SEM. The liquid is open to a water vapor ambient and the electron beam is scanned over the region where deposition is desired.

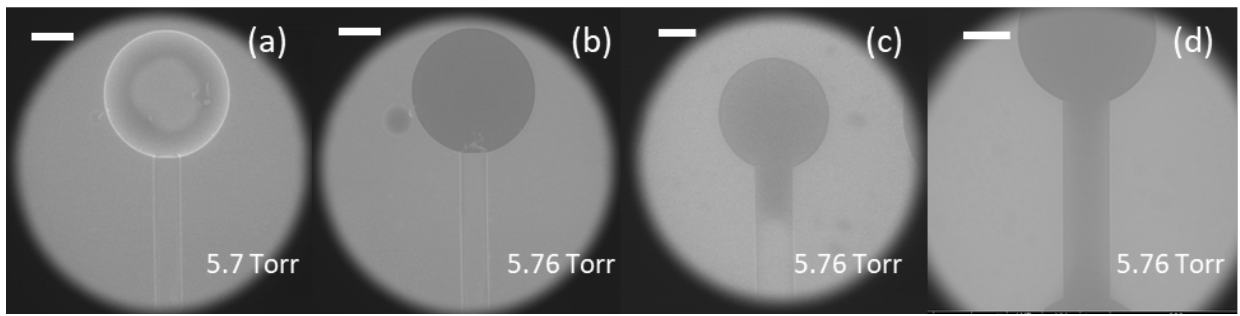

Figure 2 In-situ hydration and flow from the reservoir via the channel toward the second well. (a) After pumping the chamber to 5.7 torr, the liquid is visible with some precipitated $CuSO_4$, (b)-(d) further adjustment in the chamber pressure facilitates complete dissolution and liquid flow. Once the liquid film becomes flat, it is consistently darker than the surrounding substrate in the micrograph. The scale bars represent 100 µm.

## 2.2 Deposition

The precursor solutions used in these experiments are 125 mM, 63 mM and 31 mM CuSO₄ with 100 mM H₂SO₄ and Triton X-100 to improve wetting and liquid flow.

Figure 3 shows an in-situ image of the liquid flow and deposited patterns in the channel (a), and tilted view of the arbitrary patterns deposited with a dose of 5 µC/cm, 10 µC/cm and 15 µC/cm. Controlling the liquid thickness results in a very small amount of unintended deposition. Again, at two lower doses, we can see the trace of two lines deposited. One is on the substrate, and the other appears to have formed inside solution and merged after dehydration. At the highest line dose, it appears one solid high aspect ratio line was deposited. However, it is likely that two lines formed first and merged during the exposure. The $Cu^{2+}$ ions can be directly reduced to the metallic copper atoms via hydrated electrons generated due to the radiolysis of water molecules by the primary electrons in the liquid volume. In an alternative pathway, hydronium can react with the hydrated electrons and produce hydrogen atoms. Then hydrogen can act as the reducing agent for $Cu^{2+}$ to $Cu^{1+}$ ions. $Cu^{+1}$ is reducing to Cu(s) through a disproportionation reaction. Moreover nucleation and growth of copper occurs at the solid-liquid interface due to the secondary electron and backscattered electron from the negatively charged substrate. The deposit will build up by nucleation and growth of copper atoms and clusters. Detailed analysis of the reactions to deposit copper from aqueous copper solutions and, electron beam spatial energy deposition in the liquid in vicinity of water vapor using Monte Carlo simulations are presented in [29] and [30] respectively.

Moreover, in Figure 4 it is shown that a pattern of 200 nm pitch nested lines deposited at a relatively high dose of 20 µC/cm, are merged with a flat top surface. Figure 4(a) shows the top view of the in-situ deposit right after patterning and Fig. 4(b) is the 60° tilted view of the same pattern ex-situ. There is some imperfection in the side walls of the deposit that can either be because of poor adhesion of copper and subsequeny removal during dehydration and rinsing or because the region is depleted of copper ions during patterning due to relatively long exposure time. There is flat top for this specific pattern which indicates the growth stopped right at the uniform top surface of liquid film and did not grow beyond the liquid which is another indication of hydrated electron roles in reducing the copper ions. The measured height of the deposited structures indicate that the film thickness was around 3 µm which is consistent with the etch depth of the channel and shallow well. So, using these microwells enabled us to effectively control the liquid thickness and uniformity.

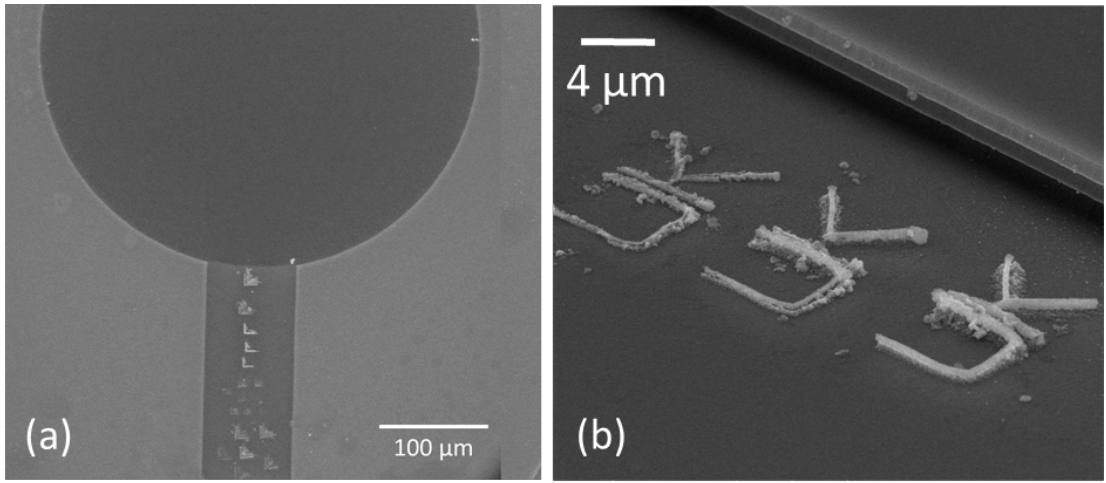

Figure 3 (a) Liquid flows from the reservoir into the channel at 5.6 torr and deposits (visible as bright, small features) can be formed. (b) Ex-situ micrograph with 60° tilt-view of an arbitrary pattern deposited in the channel with the dose of 5µC/cm, 10 µC/cm and 15 µC/cm from left to right.

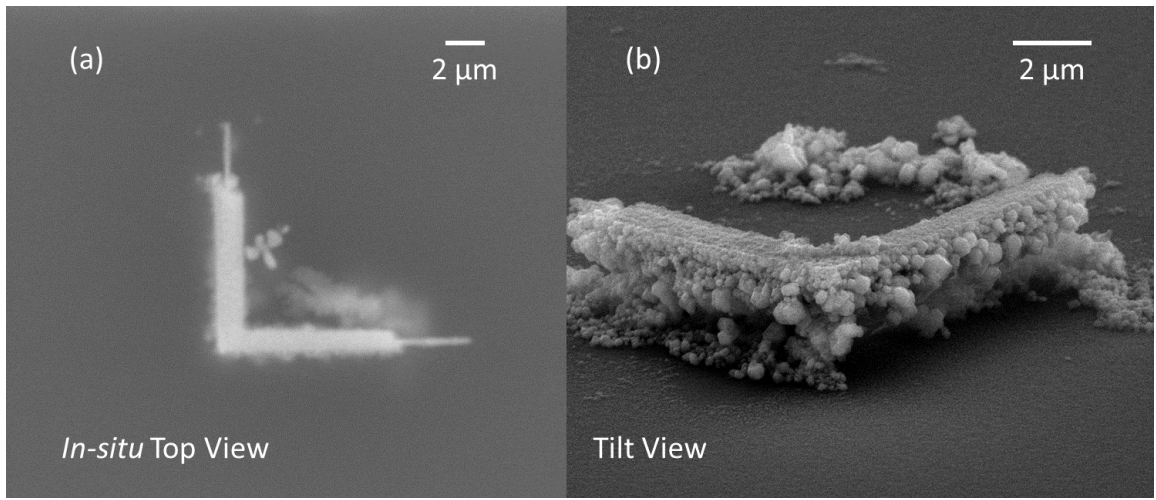

Figure 4 SEM image of 200 nm pitch nested lines deposited from acidic copper sulfate solution at a relatively high dose of 20 µC/cm, (a) *In-Situ* top view, and (b) 60° tilted view of the same pattern in the hi-vac mode.

After establishing that deposition consistently occurs, we carefully designed experiments to control and study the LP-EBID parameters. Our patterns were defined to be an array of single dots to ease the calculations for the volume of deposited copper and the efficiency of the process.

Figure 5 shows a 10 x 10 array of copper dots deposited from 63 mM (a) and 125 mM (b) $Cu^{2+}$ precursors. This pattern shows the effect of changing dwell time (dose) on each element as it varies point by point in a range of 10-100 pC/dot and 10 exposure cycles with a constant 50 ms refresh time for the cycles. As the dwell time increases, for both solutions, a larger volume of deposit is obtained. Based on our previous work [29], it is expected that the size of the deposits is controlled by the electron dose and will increase as a function of increasing electron dose or equivalently irradiation time. For both precursors, the first few copper dots are missing or displaced due to their small volume and/or poor adhesion to the substrate. Also, collateral deposition was observed at higher doses. This experiment helped us to determine the appropriate electron dose with minimal collateral deposition and adequate adhesion to study the parameters of interest, described in the next section.

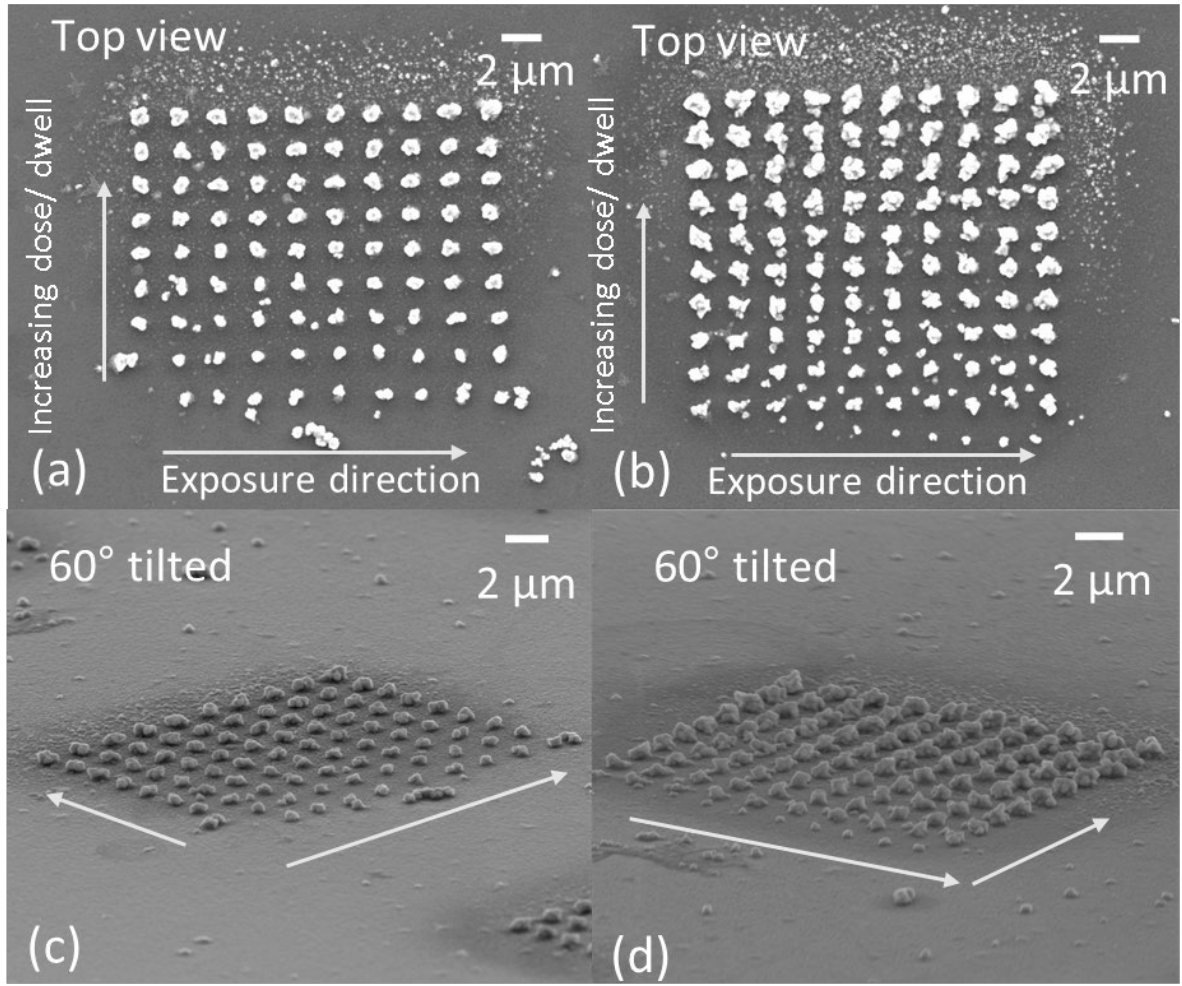

Figure 5 Example experiments used to study the deposition of copper across a wide parameter space. 10 x 10 arrays of copper dots were deposited from (a, c) 63 mM and (b, d) 125 mM $CuSO_4$: 100 mM $H_2SO_4$: 1 mM Triton X-100, with an electron dose ranging from 10 to 100 pC /dot with 10 exposure cycles and 50 ms refresh time. A larger volume of metallic Cu is deposited at higher $Cu^{2+}$ concentrations.

## 3. Results and Discussion

Our goal was to identify different deposition regimes and build a foundation for modelling. Large (~ 500 nm) arrays of dots were deposited in the microchannel with 3 μm thick liquid. We measured the volume of the deposits and inferred the deposition efficiency. The efficiency is defined as the number of copper atoms deposited per primary electrons (PE). Considering the high purity of copper deposit using liquid phase EBID of the acidified copper sulfate precursor [29] bulk density of copper was used for the calculations. ImageJ graphical tool was used to estimate the volume of copper deposits based of the geometrical shapes. Specifically, the base and the height of deposit were measured, and a conical shape was assumed in the calculations. For a few of the copper deposits the geometry was closer to cylindrical than conical, and thus a cylinder was considered to estimate the volume . To increase the accuracy of the calculations large number (a range of 25 to 75) of copper features deposited with the same experimental parameters were analysed and the mean value was used in the efficiency calculations. Parameters that have been studied in the next sections are dwell time, refresh time, cycles/total dose, and concentration.

### 3.1 Effect of dwell time and refresh time on deposition efficiency

Figure 6 (a) plots the average deposition efficiency versus refresh time from 25 patterns deposited from 125 mM $CuSO_4$:100 mM $H_2SO_4$:1 mM Triton X-100. Blue line represents 100 pC/cycle (140 ms dwell time) × 10 cycles per dot, and the red line represents 200 pC/cycle (280 ms dwell time) × 5 cycles. Thus, each dot was deposited with a 1000 pC electron dose. It is seen that the deposition efficiency depends strongly on dwell time and moderately on refresh time. Longer dwell times resulted in almost twice as efficient deposition at the same dose. While increasing the refresh time also increases efficiency, it is not as dramatic an effect.

We extend the study of the dwell time effect by plotting the number of deposited copper atoms per electron pulse as function of refresh time for the same patterns in Figure 6 (b). Interestingly, the two results are almost identical for the two different dwell times. This indicates that doubling the number of electrons in a pulse has no effect on the amount of deposited copper. Thus, the process is extremely mass-transport limited at these dwell times. $Cu^{2+}$ ions must be completed depleted in the region near the beam impact point at some dwell time lower than 140 ms. However, all the refresh times studied here allowed sufficient replenishment of $Cu^{2+}$ such that deposition continued for subsequent pulses. However, the modest increase in deposition with refresh time indicates that even 400 ms does not completely replenish the reactant in the vicinity of the beam.

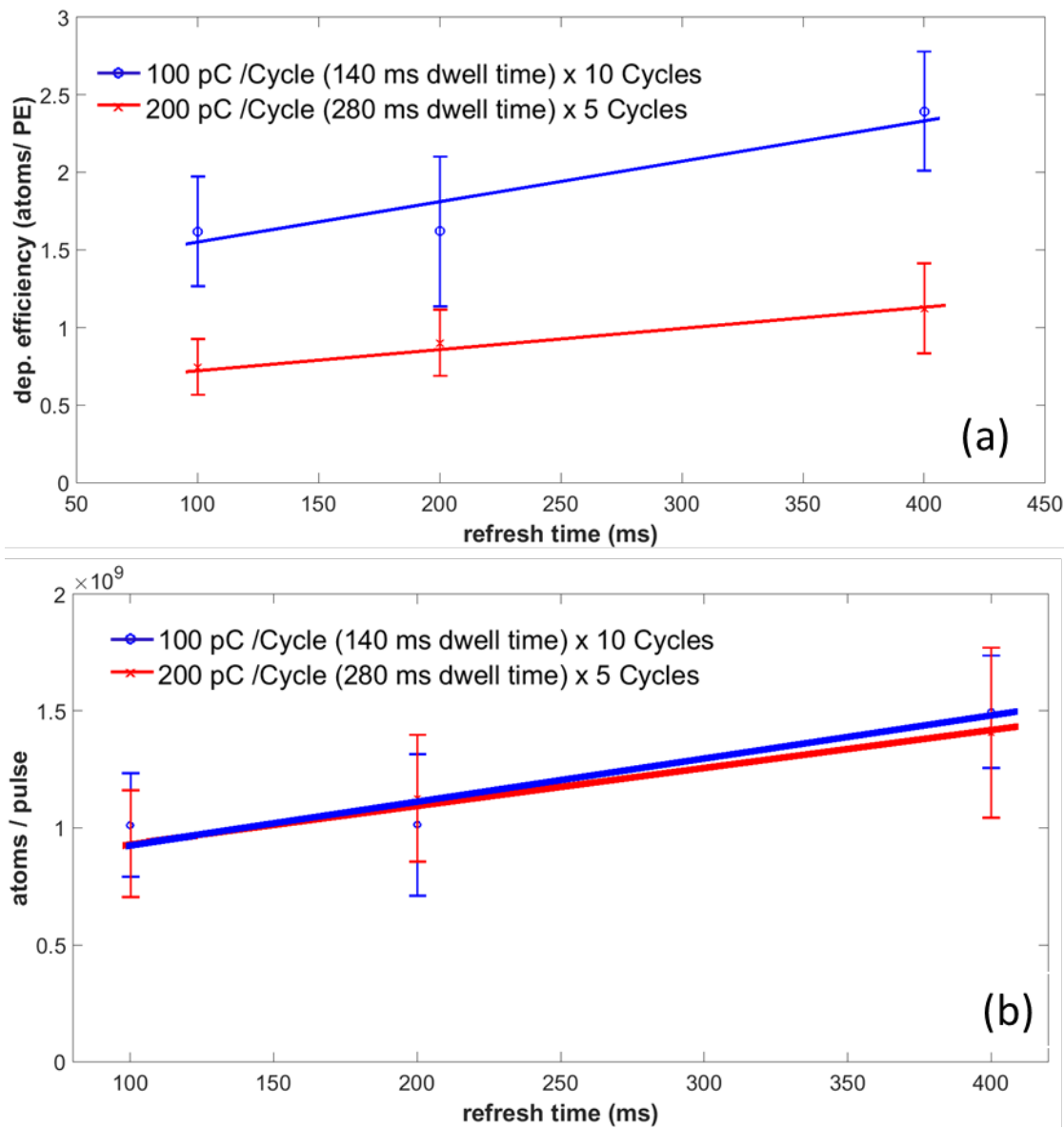

Figure 6 (a) Measured deposition efficiency in Cu atoms per primary electron (PE) for 140 ms and 280 ms dwell time per dot, and (b) measured copper atoms deposited per electron pulse for 140 ms and 280 ms dwell time per dot vs refresh time. Dot arrays were deposited from 125 mM $CuSO_4$:100 mM $H_2SO_4$:1 mM Triton X-100.

In the next step, deposition efficiency versus refresh time was measured for 20 pulses with 10 pC/pulse/dot and 15 pC/pulse/dot. The former dose corresponded to the 14 ms of dwell time and the latter to 21ms of dwell time. A solution of 125 mM $CuSO_4$:100 mM $H_2SO_4$:1 mM Triton X-100 was used. Again, three refresh times were examined for each pattern.

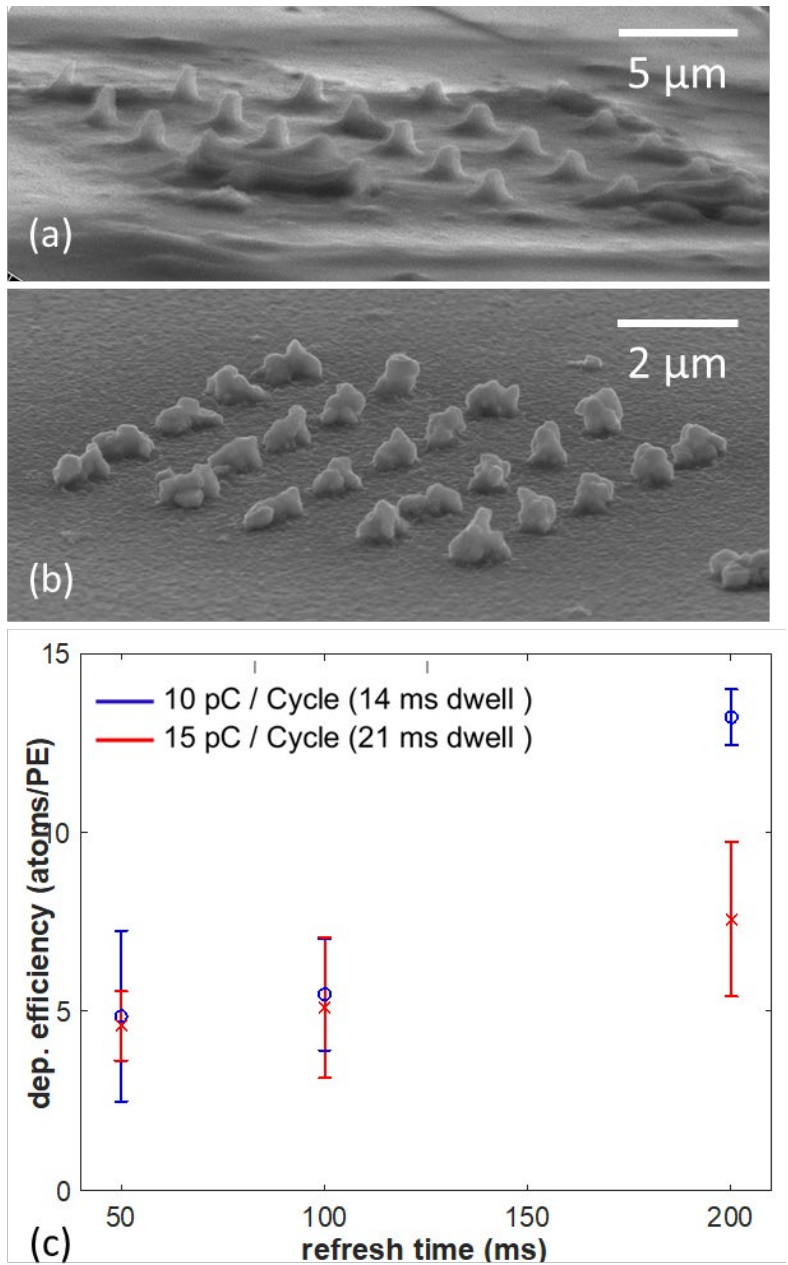

Figure 7 SEM micrographs show the transition from (a) conical to (b) granular copper deposits at the shortest and longest dwell and refresh times respectively. (c) Measured efficiency for 14 ms (blue) and 21 ms (red) dwell time per dot for dot arrays deposited from 125 mM $CuSO_4$:100 mM $H_2SO_4$:1 mM Triton X-100.

In this case, the process is neither dwell time nor refresh dependent except for 200 ms refresh. This result indicates that the process was electron limited at the specified parameters. Moreover, we observed a transition from almost conical deposits to granular deposits at long refresh times. Therefore, an aggregation mechanism vs. single point nucleation and growth governs the deposition. Finally, the high deposition efficiency of these experiments ($\approx$ 5 to 13 Cu atoms per primary electron) supports a radiation chemical mechanism for deposition [29]. In this model, each primary electron generates a large number of reducing species via radiolysis, likely atomic hydrogen in the acidic solution used here. These radicals reduce $Cu^{2+}$(aq) to Cu(s) to enable the deposition process.

### 3.2 Concentration Effect on Copper Deposition

The influence of $Cu^{2+}$ concentration was also investigated by depositing copper patterns with 31 mM and 63 mM $CuSO_4$ in the precursor. For 63 mM concentration, any refresh time longer than 200 ms resulted in extensive unintended deposition due to formation of metallic copper suspended in the liquid. Figure 8 (a) and (b) show tilted view micrographs of copper patterns deposited from low (31 mM $CuSO_4$) and high (63 mM $CuSO_4$) concentration solutions respectively; panel (a) demonstrates how collateral deposition started to effect the patterns at longer refresh time and (b) shows how long refresh resulted in granular deposits instead of conical deposits. Figure 8 (c) shows an example of high collateral deposition for 400 ms refresh time with 63 mM $CuSO_4$. We observed that at low concentrations, the process transitions from electron limited to mass transport limited as a result of changing refresh time to a longer value. Moreover, comparing the two concentrations revealed regimes where there was no nucleation or no adhesion as well as regimes with large amounts of collateral deposition. Also, the refresh dependent regime at lower $Cu^{2+}$ concentrations indicates a transition from electron limited to mass transport limited behaviour. This region also reveals practical limits on nucleation/adhesion. Figure 9 illustrates the measured deposition efficiency vs refresh time for 5×5 dot arrays deposited from 63 mM (blue), and 31 mM (red) output of this study.

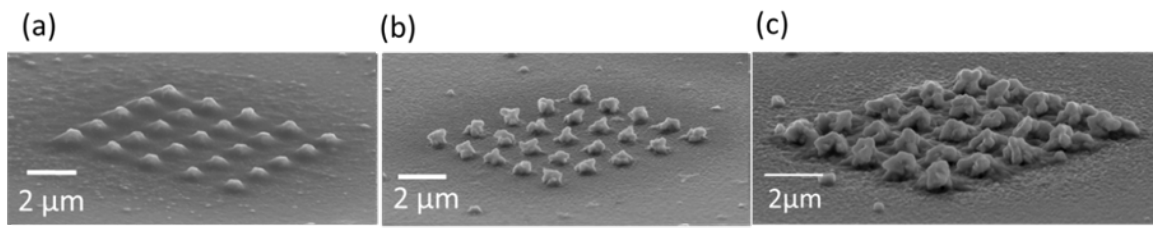

Figure 8 60° tilted SEM micrographs of copper patterns deposited from (a) 31 mM $CuSO_4$:100 mM $H_2SO_4$: 1 mM Triton X-100 at 600 ms refresh time, (b) 63 mM $CuSO_4$:100 mM $H_2SO_4$: 1 mM Triton X-100 at 200 ms refresh time, and (c) 63 mM $CuSO_4$:100 mM $H_2SO_4$: 1 mM Triton X-100 at 400 ms refresh time.

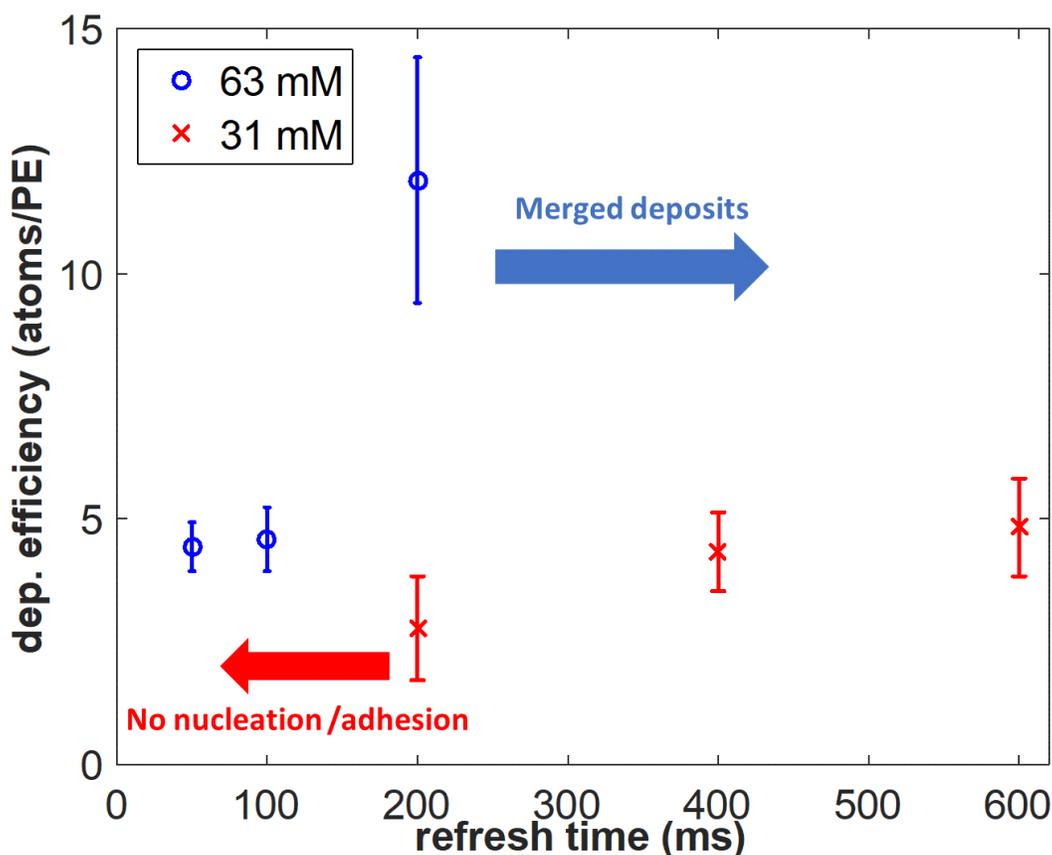

Figure 9 Measured deposition efficiency vs refresh time for 5×5 dot arrays deposited from 63 mM (blue), and 31 mM (red) $CuSO_4$:100 mM $H_2SO_4$: 1 mM Triton X-100. An electron dose of 40 pC/cycle/dot was applied in 20 cycles of exposure (total dose of 800 pC/dot). The refresh time-dependent regime at lower concentrations provides evidence of a transition from electron limited to mass transport limited deposition. Regions are also noted for which there is no nuclations or adhesion of deposits (31 mM) and merged deposits (63 mM).

## 4. Conclusions

In conclusion, micro-wells and channels enabled controlled liquid thickness and concentration, allowing us to quantify deposition efficiency. The deposition efficiency is a useful figure of merit to study the effect of deposition parameters such as concentration, dwell time, and refresh time. We established the conditions under which the process transitions from electron limited to mass transport limited. This transition had not been previously observed for LP-EBID. Moreover, an additional aggregation-based regime and high aspect ratio regime were identified throughout our studies. We observed remarkable deposition efficiency under all conditions (1-10 Cu atoms/primary electron) consistent with a radiation chemical model of deposition.

**Conflicts of interest**

There are no conflicts to declare.

**Acknowledgements**

This material is based upon work supported by the National Science Foundation under Grant No. CMMI-1538650. This work was performed in part at the University of Kentucky Center for Nanoscale Science and Engineering, the University of Kentucky Electron Microscopy Center, and the University of Louisville Micro and Nano Technology Center, members of the National Nanotechnology Coordinated Infrastructure (NNCI), which is supported by the National Science Foundation (ECCS-2025075). This work used equipment supported by National Science Foundation Grant No. CMMI-1125998.